\documentclass[11pt,twoside,superscriptaddress,showpacs]{revtex4}
\usepackage{graphicx,latexsym,amssymb,amsmath,color, multirow, mathrsfs, booktabs}

\usepackage[section]{placeins}

\newcommand{\beq}{\begin{equation}}
\newcommand{\eeq}{\end{equation}}
\newcommand{\beqn}{\begin{eqnarray}}
\newcommand{\eeqn}{\end{eqnarray}}
\newcommand{\bea}{\begin{array}}
\newcommand{\eea}{\end{array}}
\newcommand{\bsub}{\begin{subequations}}
\newcommand{\esub}{\end{subequations}}
\newcommand{\bpm}{\begin{pmatrix}}
\newcommand{\epm}{\end{pmatrix}}

\newcommand{\cals}[1]{{\mathcal #1}}
\newcommand{\scr}[1]{{\mathscr #1}}
\newcommand{\ff}[1]{\frac{1}{#1}}
\newcommand{\lc}{\left<}
\newcommand{\rc}{\right>}
\newcommand{\lr}{\left|}
\newcommand{\rl}{\right|}

\newcommand{\lrb}[1]{\left(#1\right)}
\newcommand{\lrs}[1]{\left[#1\right]}
\newcommand{\Lb}{\left\{}
\newcommand{\Rb}{\right\}}

\newcommand{\dbar}[1]{\bar{\bar{#1}}}
\newcommand{\svec}[1]{{\mbox{\boldmath${ #1}$}}}
\newcommand{\ivec}{\vec}

\newcommand{\nr}{{\text{NR}}}
\newcommand{\er}{{\text{R}}}
\newcommand{\appref}[1]{{\sf\bfseries Appendix \ref{#1}}}
\newcommand{\secref}[1]{{\sf\bfseries Section \ref{#1}}}
\newcommand{\tabref}[1]{{\sf\bfseries Table \ref{#1}}}

\newcommand{\mev}{\text{MeV}}
\newcommand{\cm}{{\text{c.m.}}}

\begin{document}

\title{Relativistic Hartree-Fock theory. Part I: density-dependent effective Lagrangians }
\author{Wen Hui Long} \email{whlong@pku.org.cn}
 \affiliation{School of Physics, Peking University, 100871 Beijing, China}
 \affiliation{CNRS-IN2P3, UMR 8608, F-91406 Orsay Cedex, France}
 \affiliation{Univ Paris-Sud, F-91405 Orsay, France}
\author{Nguyen Van Giai}
 \affiliation{CNRS-IN2P3, UMR 8608, F-91406 Orsay Cedex, France}
 \affiliation{Univ Paris-Sud, F-91405 Orsay, France}
\author{Jie Meng}
 \affiliation{School of Physics, Peking University, 100871 Beijing, China}
 \affiliation{Institute of Theoretical Physics, Chinese Academy of Sciences, Beijing, China}
 \affiliation{Center of Theoretical Nuclear Physics, National Laboratory of Heavy Ion Accelerator, 730000 Lanzhou, China}

 \begin{abstract}
Effective Lagrangians suitable for a relativistic Hartree-Fock description of nuclear systems are
presented. They include the 4 effective mesons $\sigma, \omega, \rho$ and $\pi$ with
density-dependent meson-nucleon couplings. The criteria for determining the model parameters are
the reproduction of the binding energies in a number of selected nuclei, and the bulk properties of
nuclear matter (saturation point, compression modulus, symmetry energy). An excellent description
of nuclear binding energies and radii is achieved for a range of nuclei encompassing light and
heavy systems. The predictions of the present approach compare favorably with those of existing
relativistic mean field models, with the advantage of incorporating the effects of pion-nucleon
coupling.
 \end{abstract}
\maketitle 
\section{Introduction}
The most actively investigated fields in present days nuclear
structure and reaction studies are the production and exploration of
isotopes with extreme neutron-to-proton number ratios, the so-called
exotic nuclei \cite{Mueller:1993, Tan:1995, Hansen:1995,
Casten:2000, Mul:2001, Jonson:2004, Jensen:2004}. In the last 20
years, technological breakthroughs in producing radioactive nuclear
beams have opened up entirely new and exciting frontiers for nuclear
physics \cite{Schiffer:1999}. With the development of unstable
nuclear beams \cite{Tan:1995, Mul:2001}, a number of unexpected
phenomena have been discovered like neutron and proton skins and
halos \cite{Tan:1985,Tan:1988,Warn:1995, Chul:1996}, modifications
of shell closures \cite{Ozawa:2000}, soft excitation modes
\cite{Koba:1989, Koba:1992}, the enhancement of fusion cross
sections induced by the extended matter distributions
\cite{Alm:1995,Yosh:1995}, etc. With further developments, many
other new features are likely to be found.

In the universe, novae and supernova explosions leading to neutron stars, X-ray and gamma-ray
bursts, all depend upon reactions involving nuclei that do not naturally occur on Earth. There
still remain many mysteries in the origin of elements \cite{Arnould:2001}, with the rapid
neutron-capture process accounting for the formation of about half of $A>60$ stable nuclei in
nature. The $r$-process is the most complex nucleosynthesis process from the astrophysics as well
as nuclear physics point of view \cite{Goriely:2004, Wanajo:2004}. Understanding the time scales
and energies for such processes, and hence the stellar evolution itself, should be connected with
the understanding of the exotic nuclear systems, either in astrophysical conditions or in the
extreme nuclear environment. The exotic nuclei thus provide also a new testing standard for our
understanding of phenomena of astrophysical interest.

Many of the above physics issues can be successfully studied in the framework of self-consistent
mean field approaches. Non-relativistic Hartree-Fock approaches based on Skyrme type forces
\cite{Vautherin:1972, Beiner:1975}, local density approximations of the Brueckner G-matrix
\cite{Negele:1970, Campi:1972}, and effective  Gogny type forces \cite{Decharge:1980} have been
developed. A quantitative understanding of nuclear matter as well as finite nuclei was thus
obtained. In parallel, the relativistic Brueckner-Hartree-Fock approach was applied to nuclear
matter, at first without full self-consistency \cite{Anastasio:1983}, and then fully
self-consistent calculations were carried out by Brockmann and Machleidt
\cite{Brockmann:1984,Brockmann:1990}, and by Ter Haar and Malfliet \cite{Haar:1987}. Based on a
one-boson-exchange interaction and combined with relativistic Brueckner-Hartree-Fock approach, the
relativistic approach offers an appealing framework to investigate and explore macroscopic
properties of hot and dense nuclear matter \cite{Haar:1987}. Using a density-dependent
parametrization of the Dirac-Brueckner $G$-matrix in nuclear matter, an effective
Dirac-Brueckner-Hartree-Fock model for finite nuclei was achieved by Boersma and Malfliet
\cite{Boersma:1994I, Boersma:1994II}, which also offers a wide range of applications in the nuclear
physics and astrophysics domains.

During the past years, the relativistic mean field (RMF) theory \cite{Miller:1972, Walecka:1974} has received wide
attention due to its successful description of numerous nuclear phenomena \cite{Serot:1986, Reinhard:1989, Ring:1996,
Serot:1997, Bender:2003,Meng:2006}. The RMF can be viewed as a relativistic Hartree approach with a no-sea
approximation. In the framework of the RMF, the nucleons interact via the exchanges of effective mesons, and photons.
For nuclear structure applications, the relevance of relativity is not the need for relativistic kinematics but that of
a covariant formulation which maintains the distinction between scalar and vector fields (more precisely, the zeroth
component of Lorentz four-vectors). The representations with large scalar and vector fields in nuclei, of the order of
a few hundred MeV, provide simpler and more efficient descriptions than non-relativistic approaches~\cite{Furnstahl00a,
Furnstahl00b, Furnstahl04}. The dominant evidence is the spin-orbit splitting. Other evidence includes the density
dependence of the optical potential, the observation of approximate pseudo-spin symmetry, etc. With a very limited
number of parameters adjusted to reproduce selected empirical observables, the RMF theory can describe very
satisfactorily general properties such as the masses and radii of nuclei in the whole periodic table, or specific
aspects like the observed kink in the isotopic shifts of the Pb-region \cite{Sharma:1993b}. It gives naturally the
spin-orbit potential, the origin of the pseudo-spin symmetry \cite{Arima:1969, Hecht:1969} as a relativistic symmetry
\cite{Ginocchio97,Meng98r,Meng99prc}, the spin symmetry in the anti-nucleon spectrum \cite{Zhou03prl}, etc. It also
performs quite well for nuclei far away from the line of $\beta$-stability with proper treatment of the pairing
correlation and continuum effects \cite{Meng:2006,Meng:1996,Meng:1998a, Meng:1998PRL}.

In spite of the success, there are still some shortcomings in the RMF theory. A too small value of
the symmetry energy would be obtained if the coupling constant $g_\rho$ takes on its experimental
value. Actually, important contributions due to the exchange terms are missing in the model. Among
the effective mesons which mediate the nucleon interaction, the pion meson might be the most
important one. Because of the limit of the approximation itself, the one pion exchange cannot
contribute within the framework of the RMF if parity is to be conserved. The RMF cannot take into
account the effective spin-spin interaction coming from the nucleon density through the
anti-symmetrization of the wave function. This effect is more important in spin-unsaturated systems
\cite{Marcos:1991}.

Early attempts to investigate the structure of nuclear matter and finite nuclei in the relativistic Hartree-Fock (RHF)
approximation have been made {\cite{Bouyssy:1985,Bouyssy:1987}}. The effective Lagrangians did not include density
dependence or meson self-couplings. They gave a satisfactory description of the charge densities of magic nuclei but
failed on the incompressibility. The nuclei were not bound enough, this defect being more important for light nuclei.
By introducing additional density-dependent terms in the Lagrangian density, better agreement with the experiment and
significant improvement on the incompressibility were obtained \cite{Bernardos:1993}. The density dependence was
simulated by the nonlinear self-couplings of the $\sigma$-field, but chiral symmetry is not conserved. To recover the
chiral symmetry, the non-linear self-interaction (NLSI) of the scalar field with zero-range was introduced in Ref.
\cite{Marcos:2004}. With the zero-range limit, the nonlinear self-couplings are expressed in terms of the products of
six and eight nucleon spinors, where the exchange contributions can be evaluated by the Fierz transformation. It is
shown that, in the RHF formalism the explicitly isovector-independent NLSI generates strongly density-dependent scalar,
vector and tensor nucleon self-energies. The description of nuclei thus obtained is in reasonable agreement with
experimental data.

Although some progress have been obtained in the relativistic Hartree-Fock description of nuclear
systems, it is still a long way compared to the RMF, either on the quantitative descriptions of
nuclear systems or their extrapolation to the exotic region. Especially compared with the wide
applications of RMF, the disparity seems more remarkable. Of course, the technical and numerical
difficulties are increased strongly with the inclusion of the exchange terms. On the other hand,
there is still no appropriate effective Lagrangian for the RHF approach, which does not break the
chiral symmetry nor increase the complexity of the theory itself.

{Based on the above considerations, we have constructed a density-dependent relativistic Hartree-Fock (DDRHF) theory by
quantizing a Hamiltonian which contains density-dependent meson-nucleon couplings \cite{Long:2006}. For the first time,
the RHF approach is systematically applied to effective Lagrangians with density-dependent meson-nucleon couplings. The
main finding is that, with a number of adjustable parameters comparable to that of RMF Lagrangians, one can obtain a
good quantitative description of nuclear systems without dropping the Fock terms. The pseudo-spin symmetry in finite
nuclei within our DDRHF model has been investigated in Ref. \cite{Long:2006PS}. In this paper we present the method for
obtaining the DDRHF effective Lagrangians and the results for the selected nuclei. Systematical applications of the
DDRHF in nuclear matter and finite nuclei will be shown in a forthcoming paper \cite{Long:2006b}.}

{This paper is organized as follows. The derivation of the DDRHF is briefly introduced in \secref{sec:Theory}. In
\secref{sec:Eff} are given the parametrizations of DDRHF and some initial applications in nuclear matter and in the set
of selected nuclei. A short summary is given in \secref{sec:CP}. The calculations of center-of-mass corrections for
energies, radii and charge densities are explained in the Appendix.}

\section{Theory Framework}\label{sec:Theory}
\subsection{General Formalism}
The general framework can be found in several papers published in the literature. We follow closely the formalism and
notations of {Refs. \cite{Bouyssy:1985,Bouyssy:1987}}. {The detailed expressions for nuclear matter and finite nuclei
in the spherical symmetry approximation were already given in these two references, except that they were established
for the case of density-independent coupling constants.} Therefore, we recall here only {the main steps and we briefly
explain how to treat the additional rearrangement terms brought about by our density-dependent Lagrangian.}

The starting point is an effective Lagrangian density $\cals L (x)$ constructed from the degrees of
freedom associated with the nucleon ($\psi$), two isoscalar mesons ($\sigma$ and $\omega$), two
isovector mesons ($\pi$ and $\rho$) and the photon ($A$) fields. The stationarity condition of the
action integral $\int d^4x \cals L(x)$ variations of the physical fields $\phi$ ($\phi = \psi,
\sigma, \omega, \rho, \pi$ and $A$) leads to the Euler-Lagrange equations, from which one can
deduce the equations of motion for the meson, photon and nucleon fields \cite{Bouyssy:1987}. The
meson and photon fields obey inhomogeneous Klein-Gordon equations and a Proca equation with source
terms, respectively, whereas the nucleon field obeys a Dirac equation.

The Hamiltonian operator, i.e., the $(00)$ component of the energy-momentum tensor $T^{\mu\nu}\equiv \frac{\partial\scr
L}{\partial(\partial_\mu\phi_i)}\partial^\nu\phi_i - g^{\mu\nu}\scr L$,  can be formally obtained through the general
Legendre transformation
 \beq
\scr H = T^{00} = \frac{\partial\scr L}{\partial
\dot\phi_i}\dot\phi_i - \scr L~.
 \eeq

The next step is to use the Klein-Gordon and Proca equations to formally express the meson and photon fields in terms
of nucleon fields $\psi$ only. Thus, one would be able to obtain the Hamiltonian in nucleon space in a form suitable
for studies of nuclear systems. We make the simplifying assumption of neglecting the time component of the four-momenta
carried by the mesons, i.e., the meson fields are assumed to be time independent. This amounts to neglect the
retardation effects. The energies involved are small compared to the masses of the exchanged mesons, so that this
approximation should be valid for the $\sigma$-, $\omega$- and $\rho$-induced interactions, and also, to a lesser
extent for the pion. Then, the meson propagators have the usual Yukawa form,
 \beq\label{Yukawaform}
{D_\sigma(\svec x_1,\svec x_2) = \frac{1}{4\pi}
\frac{e^{-m_\sigma\lr\svec x_1-\svec x_2\rl}}{\lr\svec x_1-\svec
x_2\rl},}
 \eeq
and the nucleonic Hamiltonian can be written as
 \beq\label{Hamiltonian}
H = \int d^3x\lrs{\bar\psi\lrs{-i\svec\gamma\cdot\svec\nabla + M}\psi} +\ff2 \int d^3x_1 d^4 x_2
\sum_{{i=\sigma,\omega},\atop\rho,\pi, A}\bar\psi(x_1) \bar\psi(x_2) \Gamma_i(1,2) D_i(x_1,x_2) \psi(x_2)\psi(x_1),
 \eeq
{where the interaction vertices are generalizations of those of Ref. \cite{Bouyssy:1987}:}
 \bsub\begin{align}
\Gamma_\sigma(1,2)\equiv& -g_\sigma(1) g_\sigma(2),\\
\Gamma_\omega(1,2)\equiv& +g_\omega(1)\gamma_\mu(1) g_\omega(2)\gamma^\mu(2),\\
\Gamma_\rho(1,2)\equiv& +g_\rho(1)\gamma_\mu(1)\ivec\tau(1)\cdot
g_\rho(2)\gamma^\mu(2)\ivec\tau(2),\\ \Gamma_\pi(1,2)\equiv&
-\lrs{\textstyle{\frac{f_\pi}{m_\pi}}\ivec\tau\gamma_5\gamma_\mu\partial^\mu
}_1\cdot\lrs{\textstyle{\frac{f_\pi}{m_\pi}}\ivec\tau\gamma_5\gamma_\nu\partial^\nu }_2,\\
\Gamma_A(1,2)\equiv& +\frac{e^2}{4}\lrs{ \gamma_\mu(1-\tau_3)}_1 \lrs{ \gamma^\mu(1-\tau_3)}_2.
 \end{align}\esub

The nucleon field operators $\psi$ and $\psi^\dag$ are expanded as \cite{Bouyssy:1987}:
 \bsub\label{expansions}\begin{align}
\psi(x) =& \sum_i \lrs{ f_i(\svec x) e^{-i \varepsilon_i t} c_i +
g_i(\svec x) e^{i\varepsilon_i' t} d_i^\dag},\\
\psi^\dag(x) =& \sum_i \lrs{ f_i^\dag(\svec x) e^{i \varepsilon_i t} c_i^\dag + g_i^\dag(\svec x)
e^{-i\varepsilon_i' t} d_i},
 \end{align}\esub
where $f_i(\svec x)$ and $g_i(\svec x)$ are complete sets of Dirac spinors, $c_i$ and $c_i^\dag$
represent annihilation and creation operators for nucleons in a state $i$, while $d_i$ and
$d_i^\dag$ are the corresponding ones for antinucleons. Since we will restrict ourselves to the
Hartree-Fock approximation level and study the exchange corrections to the mean field approach, we
keep the same level of approximation, the so-called no-sea approximation, i.e., the $d$ and
$d^\dag$ terms are omitted in the expansions. Their inclusion leads to divergences and requires a
cumbersome renormalization procedure \cite{Serot:1986}. The $\{f_i\}$ states will be determined by
the self-consistent HF scheme.

The trial ground state $\lr\Phi_0\rc$ is the HF Slater determinant built on the lowest $\{f_i\}$ states. The energy
functional, i.e., the expectation value of the Hamiltonian can be obtained as
 \beq\label{EnergyF}
E \equiv \lc\Phi_0\rl H\lr\Phi_0\rc = \lc\Phi_0\rl T\lr\Phi_0\rc
+\sum_i \lc \Phi_0\rl V_i\lr\Phi_0\rc.
 \eeq
For the two-body interaction parts $\lc \Phi_0\rl V_i\lr\Phi_0\rc$,
one obtains two types of contributions, the direct (Hartree) and
exchange (Fock) terms. For example, the $\sigma$-meson contributions
are:
 \bsub\begin{align}
E_\sigma^D = &-\ff2 \int d\svec x\
g_\sigma\lc\Phi_0\rl\rho_s\lr\Phi_0\rc\int d\svec x'
g_\sigma\lc\Phi_0\rl\rho_s\lr\Phi_0\rc D_\sigma(\svec x, \svec x')\\
E_\sigma^E =& +\ff2\int d\svec x \int d\svec x' \sum_{\alpha,\beta} \lrs{ g_\sigma\bar f_\alpha
f_\beta}_{\svec x} D_\sigma(\svec x, \svec x') \lrs{ g_\sigma \bar f_\beta f_\alpha}_{\svec x'},
 \end{align}\esub
where $\rho_s$ is the scalar density. The states $\{f_\alpha\}$ are solutions of a Dirac equation containing a
self-energy term $\Sigma$. The stationarity of the energy (\ref{EnergyF}) with respect to variations of the
$\{f_\alpha\}$ determines self-consistently $\Sigma$.

To be complete, we would like to mention about the contributions from the one-pion exchange. At the
Hartree-Fock level, it is well known  from general arguments (chiral symmetry) and from the
construction of the NN potential (with pair suppression mechanism) that one should adopt the
pseudo-vector coupling to obtain reasonable results in the one-pion exchange approximation. Here,
we use the pseudo-vector coupling for the pion, as it is done in Ref. \cite{Bouyssy:1987}. At the
Hartree (RMF) level, the contributions from the one-pion exchange are reduced to zero if the
variational space respects parity conservation while in RHF these contributions appear naturally.

In the recent years, the RMF with density-dependent meson-nucleon couplings (DDRMF) has been
actively developed \cite{Brockmann:1992, Lenske:1995, Fuchs:1995, Typel:1999, Niksic:2002, Long04}.
In this work, the density-dependent relativistic Hartree-Fock (DDRHF) theory is derived by jointly
using the RHF approach with an effective Lagrangian whose meson-nucleon couplings are
density-dependent. Similarly to the DDRMF, the coupling constants $g_\sigma$, $g_\omega$, $g_\rho$
and $f_\pi$ are taken as functions of the baryonic density $\rho_b$, the zeroth component of the
nucleon current $j^\nu = \bar\psi\gamma^\nu\psi$. For the $\sigma$- and $\omega$-mesons, the
density-dependent behaviors of $g_\sigma$ and $g_\omega$ are chosen as
 \beq
g_i(\rho_b) = g_i(\rho_0) f_i(\xi)~~~~~~~~\text{ for } i =\sigma,
\omega
 \eeq
where
 \beq
f_i(\xi ) = a_i\frac{1 + b_i(\xi +d_i)^2}{1+ c_i(\xi +d_i)^2},
 \eeq
is a function of $\xi =\rho_b/\rho_0$, and $\rho_0$ denotes the saturation density of symmetric
nuclear matter. For the functions $f_i(x)$, five constraint conditions $f_i(1) = 1, f_\sigma''(1) =
f_\omega''(1)$ and $f_i''(0) =0$ are introduced to reduce the number of free parameters.

For $g_\rho$ and $f_\pi$, an exponential density-dependence is adopted
 \bsub\beqn
g_\rho(\rho_b) & =& g_\rho(0) e^{-a_\rho\xi}, \\
f_\pi(\rho_b)  & =& f_\pi(0) e^{-a_\pi\xi}.
 \eeqn\esub
Because of the energy-momentum conservation condition, the density-dependence in meson-nucleon
couplings will lead to a new term in the HF self-energy, the rearrangement term $\Sigma_R^\mu$.
Thus, the total HF self-energy is a sum of a Hartree (direct) term $\Sigma_D$, Fock (exchange) term
$\Sigma_F$ and rearrangement term $\Sigma_R$.

\subsection{Nuclear Matter}
In nuclear matter, the self-energy $\Sigma$ can be written quite generally as
 \beq\label{self-energy}
\Sigma(p,
p_F)=\Sigma_S(p,p_F)+\gamma_0\Sigma_0(p,p_F)+\svec\gamma\centerdot
\hat{\svec p}\Sigma_V(p,p_F),
 \eeq
where $\hat{\svec p}$ is the unit vector along $\svec p$ and $p_F$ is the Fermi momentum. The
tensor piece $\gamma_0\svec\gamma\cdot\hat{\svec p}\Sigma_T(p)$ does not appear in the HF
approximation for nuclear matter. The scalar component $\Sigma_S$, time component $\Sigma_0$ and
space component $\Sigma_V$ of the vector potential are functions of $p=(E(p),\svec p)$.

With the general form of the self-energy, the Dirac equation in infinite nuclear medium can be written as
{\cite{Bouyssy:1985}}
 \beq\label{DiracNM}
\lrs{\svec\gamma\cdot\svec p^*+M_S^*} u(p,s)=\gamma_0E^* u(p,s).
 \eeq
The starred quantities are defined by
 \beq\label{starred}
\begin{split}
\svec p^*=& \svec p+\hat{\svec p} \Sigma_V(p),\\
M_S^*=&M+\Sigma_S(p),\\
E^*=&E(p)-\Sigma_0(p),
\end{split}
 \eeq
where $M_S^*$ is the scalar nucleon mass, and $E$ is the single-particle energy in the medium. {The calculation of the
self-consistent self-energies follows the method indicated in Refs. \cite{Bouyssy:1985, Bouyssy:1987}.} In DDRHF, the
density-dependence of the coupling constants gives additional {contributions to the vector self-energy $\Sigma_0$,
namely the direct ($\Sigma_R^D$) and exchange ($\Sigma_R^E$) rearrangement terms $\Sigma_R$:}
 \begin{eqnarray}
\Sigma_R &=& \Sigma_{R,\sigma}^D + \Sigma_{R,\omega}^D +
\Sigma_{R,\rho}^D + \Sigma_{R,\sigma}^E + \Sigma_{R,\omega}^E +
\Sigma_{R,\rho}^E + \Sigma_{R,\pi}^E\nonumber \\
 &\equiv& \Sigma^D_R + \Sigma^E_R~,
 \end{eqnarray}
where
 \bsub\beqn
\Sigma_{R}^D &=& \frac{1}{\rho_0}\lrs{-\frac{g_\sigma}{m_\sigma^2}\rho_s^2\frac{\partial
g_\sigma}{\partial x}+\frac{g_\omega}{m_\omega^2}\rho_b^2\frac{\partial g_\omega}{\partial x}
+\frac{g_\rho}{m_\rho^2}\rho_b^{(3)}\rho_b^{(3)}\frac{\partial g_\rho}{\partial x} },\\
\Sigma_R^E   &=& \frac{1}{\rho_0}\lrs{\frac{\delta \varepsilon_\sigma^E}{\delta
g_\sigma}\frac{\partial g_\sigma}{\partial x} + \frac{\delta \varepsilon_\omega^E}{\delta
g_\omega}\frac{\partial g_\omega}{\partial x} + \frac{\delta \varepsilon_\rho^E}{\delta
g_\rho}\frac{\partial g_\rho}{\partial x}+ \frac{\delta \varepsilon_\pi^E}{\delta
f_\pi}\frac{\partial f_\pi}{\partial x}},
 \eeqn\esub
with $x=\rho_b/\rho_0$. The rearrangement term $\Sigma_R$ can also be explicitly written in terms of self-energies. For
example, the contribution from the $\sigma$-meson reads as
 \beq
\Sigma_{R,\sigma} = \frac{\partial g_\sigma}{\partial \rho_b}\ff{g_\sigma}
\sum_\tau\ff{\pi^2}\int\lrs{\hat M(p) \Sigma_{\tau,S}^{\sigma}(p) + \Sigma_{\tau,0}^{\sigma}(p) +
\hat P(p)\Sigma_{\tau, V}^{\sigma}(p)} p^2 dp.
 \eeq

From the self-energies, the starred functions (\ref{starred}) can be determined. Finally, a self-consistent procedure
is achieved for a given baryonic density $\rho_b$ and relative neutron excess $\beta\equiv (N-Z)/A$ as,
 \beq
^0\Sigma_S,~ ^0\Sigma_V\to \Lb\bea{llll}\hat M(p)&\to&\Lb\bea{l}\rho_s\\[0.5em]\Sigma_S^E\eea\Rb\to& ^1\Sigma_S
\\[1.5em]\hat P(p)&\to&^1\Sigma_V(p)\eea\right. \to \text{New iteration}.
 \eeq

\subsection{Spherical Nuclei}
In spherical nuclei, a single-particle state with energy $E_a$ is specified by the set of quantum
numbers $\alpha = \lrb{a, m_a}$ and $a = \lrb{\tau_a, n_a, l_a, j_a}$, where $\tau_a = 1$ for
neutrons and $-1$ for protons. The Dirac spinor $f_\alpha(\svec r)$ in the expansion
(\ref{expansions}) can be written as,
 \beq\label{SpinorSP}
f_\alpha(\svec r) =\ff r \bpm iG_a (r) \\ F_a(r)\hat{\svec\sigma}
\centerdot \hat{\svec r}\epm \scr Y_{j_am_a}^{l_a}(\hat{\svec r})
\chi_{\ff2}( \tau_a),
 \eeq
where $\chi_{\ff2}(\tau_a)$ is the isospinor, $G_a$ and $F_a$ correspond to the radial parts of upper and lower
components, respectively, $\scr Y_{j_am_a}^{l_a}$ is the spinor spherical harmonics. The Dirac spinor (\ref{SpinorSP})
is normalized according to
 \beq
\int d\svec r f_\alpha^\dag (\svec r) f_\alpha(\svec r) = \int dr \lrs{ G_a^2(r) + F_a^2(r)} = 1.
 \eeq
{With the help of the Dirac spinors (\ref{SpinorSP}) the expression for the total energy (\ref{EnergyF}) can be
obtained \cite{Bouyssy:1987}.}

Then, the Hartree-Fock equations can be derived by requiring the total binding energy $E_B$
 \beq
E_B = \lc \phi_0\rl H\lr \phi_0\rc - AM = E -AM
 \eeq
to be stationary with respect to norm-conserving variations of the
spinors $f_a$ (or, of $G_a$ and $F_a$):
 \beq\label{VariationSP}
\delta\lrs{ E_B - \sum_\alpha E_a \int f_a^\dag f_a d\svec r} =0~.
 \eeq
{The Hartree-Fock equations in coordinate space take the form:}
 \bsub\label{RHF1}\beqn
E_aG_a(r) &=& -\lrs{\frac{d}{dr}-\frac{\kappa_a}{r}} F_a(r) + \lrs{
M+\Sigma_{S}(r) + \Sigma_{0}(r)} G_a(r) + Y_a(r)\label{HF1},\\
E_aF_a(r) &=& +\lrs{\frac{d}{dr}+\frac{\kappa_a}{r}} G_a(r) -\lrs{ M+\Sigma_{S}(r) - \Sigma_{0}(r)}
F_a(r) + X_a(r)\label{HF2},
 \eeqn\esub
where $\Sigma_S$ and $\Sigma_0$ contain the contributions of all mesons to the direct (Hartree) potential and the
rearrangement potential $\Sigma_R$, $X_a$ and $Y_a$ are the convolution integrals of the non-local exchange (Fock)
potentials with $F_a$ and $G_a$, respectively. The detailed expressions of the Hartree and Fock potentials can be found
in Ref. \cite{Bouyssy:1987}. {Similarly to} nuclear matter, the rearrangement potential $\Sigma_R$ due to the
density-dependence {of the meson-nucleon couplings} can be obtained from the variation of the energy functional with
respect to the baryonic density $\rho_b$. {As an example, the rearrangement potential due to the $\sigma$-N coupling}
can be written as
 \beq
\Sigma_R^{(\sigma)}(r) = \frac{\partial g_\sigma}{\partial\rho_b}\lrs{ \rho_s(r) \sigma(r) +
\ff{g_\sigma}\ff{r^2} \sum_\alpha \hat j_a^2 \lrb{G_a(r)Y_\alpha^{(\sigma)}(r)  + F_a(r)
X_\alpha^{(\sigma)}(r)}}.
 \eeq
Notice that the rearrangement potential is local.

{The system of coupled integro-differential equations (\ref{RHF1}) is rather cumbersome to solve.} It is possible to
transform it into an equivalent system of coupled differential equations, which can be solved iteratively by the
standard techniques used in the RMF approach\cite{Meng:1998a}. To this end, one can follow the method of
Ref.\cite{Bouyssy:1987} and rewrite the functions $X_a(r)$ and $Y_a(r)$ in the form:
 \bsub\begin{align}
X_\alpha(r) =& \frac{G_a(r) X_\alpha(r)}{G_a^2(r) + F_a^2(r)} G_a(r) + \frac{F_a(r) X_\alpha(r)}{G_a^2(r) + F_a^2(r)}
F_a(r) \equiv X_{\alpha, G_a}(r) G_a(r) + X_{\alpha, F_a}(r) F_a(r),\\
Y_\alpha(r) =& \frac{G_a(r) Y_\alpha(r)}{G_a^2(r) + F_a^2(r)} G_a(r) + \frac{F_a(r) Y_\alpha(r)}{G_a^2(r) + F_a^2(r)}
F_a(r) \equiv Y_{\alpha, G_a}(r) G_a(r) + Y_{\alpha, F_a}(r) F_a(r),
 \end{align}\esub
{Then, the integro-differential equations (\ref{RHF1}) are formally transformed into a homogeneous differential
system,}
 \bsub\label{RHFD}\begin{align}
E_aG_a(r) =& -\lrs{\frac{d}{dr}-\frac{\kappa_a}{r} - Y_{\alpha, F_a}(r)}
F_a(r) + \lrs{ M+\Sigma_{S}(r) + \Sigma_{0}(r) + Y_{\alpha, G_a}(r)} G_a(r),\\
E_aF_a(r) =& +\lrs{\frac{d}{dr}+\frac{\kappa_a}{r}+X_{\alpha, G_a}(r)} G_a(r) -\lrs{ M+\Sigma_{S}(r) -
\Sigma_{0}(r)-X_{\alpha, F_a}(r)} F_a(r),
 \end{align}\esub
{which has the same structure as the RMF equations. This can be solved iteratively as in the RMF approach.}

{It is very important to treat the $T=1$ pairing correlations if one aims at a quantitative description of finite
nuclei. From a mean field point of view, one should ultimately adopt a Hartree-Fock-Bogoliubov approach. At this stage
we only use the BCS approximation \cite{Lane:1964, Ring:1980} to include pairing effects, as it is widely done in
non-relativistic and relativistic mean field approaches.} The pairing matrix elements are calculated with a zero-range,
density-dependent interaction of the type \cite{Dobaczewski:1996}
 \beq
V(r_1, r_2) = V_0 \delta(r_1-r_2) \lrs{ 1- \frac{\rho_b(r)}{\rho_0}}~,
 \eeq
with $V_0$ = -850 MeV.fm$^3$ and $\rho_0$ denotes the saturation density of nuclear matter. The active pairing space is
limited to the single-particle states below +5 MeV. {The unbound states are calculated by imposing a box boundary
condition, and one checks that the overall results are not affected by the choice of the box size. The Hartree-Fock-BCS
equations are solved self-consistently by iteration just like in  RMF-BCS calculations \cite{Reinhard:1989}. This is a
standard procedure where the presence of the Fock terms does not bring any additional difficulty.}

The correction of the center-of-mass motion is treated in a microscopic way as in Ref. \cite{Long04}. The detailed
expressions are given in \appref{Sec:CMC}. For the radii, both direct and exchange corrections (see (\ref{CMC:radii}))
are taken into account to be consistent with the theory itself. To calculate the charge densities and radii we take
into account both the proton form factor and the center-of-mass corrections (see \appref{Sec:CMC}).

\section{Effective interactions}\label{sec:Eff}

The long standing problem in the RHF approach is to provide an appropriate quantitative description of finite nuclei
and nuclear matter. The present work aims at determining such effective interactions for the DDRHF. As already done
with the RMF theory \cite{Long04}, the multi-parameter fitting is performed with the Levenberg-Marquardt method
\cite{Press:1992}. In the DDRHF, 8 parameters for the density-dependence of $g_\sigma$ and $g_\omega$ are reduced to 3
with the five constraint conditions mentioned in Subsec. II.A . Among the effective mesons which mediate the
interaction, the pion meson is the most important one. Within the RMF model the $\sigma $-meson takes into account only
the two-pion exchange in a phenomenological way. In the Hartree approximation the one-pion exchange does not contribute
because of parity conservation. This approach cannot take into account the effective spin-spin interaction coming from
the nucleon density through the anti-symmetrization of the wave function. Because of its relatively small mass the pion
cannot be approximated by a zero range force with a rather trivial exchange term. In nuclear matter and in
spin-saturated systems the one-pion exchange plays only a minor role. However, it has an important influence on
single-particle properties in spin-unsaturated systems \cite{Marcos:1991}. In the DDRHF, the contributions of one-pion
exchange can be included naturally.

With similar considerations as for the case of $\rho$-N coupling in DDRMF, an exponential density-dependent behavior
for the $\pi$-N coupling is adopted here. As further constraints, we impose that the coupling constants $g_\rho(0)$ and
$f_\pi(0)$ reach their experimental values at $\rho$=0. Thus, there are in total {\bf 8} free parameters in the DDRHF:
the mass $m_\sigma$ of the $\sigma$-meson , the couplings constants $g_\sigma(\rho_0)$ , $g_\omega(\rho_0)$ and the 3
parameters describing their density dependence, and the $a_\rho$ and $a_\pi$ parameters for the density dependence of
the $\rho$-N and $\pi$-N couplings. This gives the new effective interaction PKO1 shown in \tabref{tab:RHFS}, where
the masses of the nucleon, $\omega$-, $\rho$- and $\pi$-mesons are respectively taken as $M = 938.9$MeV, $m_\omega =
783.0$MeV, $m_\rho = 769.0$MeV and $m_\pi = 138.0$MeV, and the experimental values for the coupling constants $g_\rho$
and $f_\pi$ are respectively $2.629$ and $1.0$. In order to investigate the role of one-pion exchange, two other
parameterizations (see \tabref{tab:RHFS}) are also developed as PKO2 without $\pi$-N coupling, where $g_\rho(0)$ is
free to be adjusted, and PKO3 with free $g_\rho(0)$, where $a_\pi$ is adjusted by hand.

 \begin{table}[htbp]\centering\setlength{\tabcolsep}{4pt}
\caption{The effective interactions of the DDRHF: PKO1, PKO2 and PKO3. The masses of nucleon, $\omega$-, $\rho$- and
$\pi$-mesons are respectively taken (in MeV) as $M = 938.9$, $m_\omega = 783.0, m_\rho = 769.0$ and $m_\pi =
138.0$.}\label{tab:RHFS}
 \begin{tabular}{ccccccccc}\toprule[1.5pt]\toprule[0.5pt]
&$m_\sigma$(MeV)&$g_\sigma(\rho_0)$&$g_\omega(\rho_0)$&$g_\rho(0)$&$f_\pi(0)$&$a_\rho$&$a_\pi$&$\rho_0$
\\\midrule[0.5pt]
PKO1&525.769084&8.833239&10.729933&2.629000&1.000000&0.076760&1.231976&0.151989\\
PKO2&534.461766&8.920597&10.550553&4.068299&$-$&0.631605&$-$&0.151021\\
PKO3&525.667686&8.895635&10.802690&3.832480&1.000000&0.635336&0.934122&0.153006\\
\midrule[0.5pt]\midrule[0.5pt]
&$a_\sigma$&$b_\sigma$&$c_\sigma$&$d_\sigma$&$a_\omega$&$b_\omega$&$c_\omega$&$d_\omega$
\\\midrule[0.5pt]
PKO1&1.384494&1.513190&2.296615&0.380974&1.403347&2.008719&3.046686&0.330770\\
PKO2&1.375772&2.064391&3.052417&0.330459&1.451420&3.574373&5.478373&0.246668\\
PKO3&1.244635&1.566659&2.074581&0.400843&1.245714&1.645754&2.177077&0.391293\\
\bottomrule[0.5pt]\bottomrule[1.5pt]
 \end{tabular}
 \end{table}

In these parameterizations, the masses of the nuclei $^{16}$O, $^{40}$Ca, $^{48}$Ca, $^{56}$Ni, $^{68}$Ni, $^{90}$Zr,
$^{116}$Sn, $^{132}$Sn, $^{182}$Pb, $^{194}$Pb, $^{208}$Pb and $^{214}$Pb are fitted. The bulk properties of symmetric
nuclear matter, i.e., the saturation density $\rho_0$, the compression modulus $K$ and the symmetry energy $J$ at
saturation point are also included as constraints to improve the parameterizations. These inputs have been used to
minimize the least square error,
 \beq
\chi^2 (\svec a) = \sum_{i=1}^N\lrs{\frac{ y_i^{\text{exp}} - y(x_i;\svec a)}{\sigma_i}}^2,
 \eeq
where $y_i$ and $\sigma_i$ are the observables and corresponding weights. The vector $\svec a$ is the ensemble of the
free parameters. {Notice that the radii are not included as observables in the fitting procedure. Only the masses of
the above nuclei and the bulk properties of nuclear matter are taken into account as observables. The reason is that
the radii are much less sensitive to the changes in the parameters.} For the parameter search, the best choice is to
find {a set of observables} which have more or less the same sensitivity to the parameters. {There would be no special
difficulty to include the radii in the list of fitted data, but we have not made this choice. It is thus more
remarkable that the radii calculated with our models have a good agreement with the data, as it will be seen in Table
V.}

{The properties of nuclear matter such as the saturation point, the compression modulus $K$ and the symmetry energy $J$
are just empirically determined quantities which are known with some errors and therefore, they can be slightly varied
to improve the description of finite nuclei.} In fact, it is possible to obtain fairly good description of radii by
properly choosing the values of bulk properties of nuclear matter. The isospin related properties in nuclei can also be
well described with an appropriate choice of the symmetry energy $J$. In \tabref{tab:nucm} are shown the compression
modulus $K$, the saturation baryonic density $\rho_b$, the symmetry energy $J$, the scalar (Dirac) mass $M_S^*$, the
non-relativistic and relativistic effective masses $M_\nr^*$ and $M_\er^*$ \cite{Long:2006} at the saturation point of
nuclear matter. The results are calculated by the DDRHF with PKO1, PKO2 and PKO3, the RMF with PK1, PK1R, PKDD
\cite{Long04}, NL3 \cite{Lalazissis:1997}, TW99 \cite{Typel:1999} and DD-ME1 \cite{Niksic:2002}, and the RHF approach
with set e in Ref. \cite{Bouyssy:1987}, HFSI \cite{Bernardos:1993}, ZRL1 \cite{Marcos:2004}. It is found that the
values of the compression modulus $K$, saturation density $\rho_0$, symmetry energy $J$ and the binding energy per
particle $E/A$ provided by the DDRHF approach are all in the currently acceptable regions \cite{Vretenar:2003}, namely
$K\sim\lrs{250\mev, 270\mev}, J\sim\lrs{32\mev, 36\mev}$ and $E/A \sim 16\mev$. Compared with earlier RHF applications
\cite{Bouyssy:1987}, the improvement on the compression modulus is quite significant. It is worthwhile to mention that
the values of effective mass $M_\nr^*$ predicted by the DDRHF with PKO series are quite close to the empirical value
$M^* = 0.75$ \cite{Jaminon:1989}.

 \begin{table}[htbp]\centering\setlength{\tabcolsep}{0.2cm}
\caption{ {The bulk properties of symmetric nuclear matter calculated in DDRHF with PKO1, PKO2 and PKO3. The results
predicted by the RMF theory with PK1, PKDD, NL3 and DD-ME1, and the RHF approach with set e in Ref.
\cite{Bouyssy:1987}, HFSI \cite{Bernardos:1993}, ZRL1 \cite{Marcos:2004} are also given for comparison.
}}\label{tab:nucm}
 \begin{tabular}{c|ccc|cccc}\toprule[1.5pt]\toprule[0.5pt]
&$K$(MeV)&$\rho_0$(fm$^{-3}$)&$J$(MeV) &$E/A$(MeV)& $M_S^*(p_f)/M$& $M_\nr^*(p_f)/M$&$M_\er^*/M$\\
\midrule[0.5pt]
 {PKO1} &250.24&0.1520&34.371&-15.996&{0.5900}& {0.7459}&0.7272\\
 {PKO2} &249.60&0.1510&32.492&-16.027&{0.6025}& {0.7636}&0.7447\\
 {PKO3} &262.47&0.1530&32.987&-16.041&{0.5862}& {0.7416}&0.7229\\ \midrule[0.5pt]
 {PK1}  &282.69&0.1482&37.641&-16.268&{0.6055}& {0.6811}&0.6642\\
 {PK1R} &283.67&0.1482&37.831&-16.274&{0.6052}& {0.6812}&0.6639\\
    TM1 &281.16&0.1452&36.892&-16.263&{0.6344}& {0.7074}&0.6900\\
    NL3 &271.73&0.1483&37.416&-16.250&{0.5950}& {0.6720}&0.6547\\ \midrule[0.5pt]
 {PKDD} &262.18&0.1500&36.790&-16.267&{0.5712}& {0.6507}&0.6334\\
   TW99 &240.27&0.1530&32.767&-16.247&{0.5549}& {0.6371}&0.6198\\
 DD-ME1 &244.76&0.1520&33.069&-16.202&{0.5779}& {0.6574}&0.6403\\ \midrule[0.5pt]
 Ref. \cite{Bouyssy:1987}&465.00&0.1484&28.000&-15.750& 0.5600&$-$&$-$ \\
 HFSI         &250.00&0.1400&35.000&-15.750& 0.6100&$-$&$-$ \\
 ZRL1         &250.00&0.1550&35.000&-16.390& 0.5800&$-$&$-$ \\
\bottomrule[0.5pt]\bottomrule[1.5pt]
 \end{tabular}
 \end{table}

For the description of finite nuclei, the comparisons between the DDRHF and the previous RHF approaches
\cite{Bouyssy:1987, Bernardos:1993, Marcos:2004} are shown in \tabref{tab:comparison}. It is found that the DDRHF
provides successful quantitative descriptions of binding energies, spin-orbit splittings and charge radii for the
nuclei $^{16}$O, $^{40}$Ca, $^{48}$Ca, $^{90}$Zr and $^{208}$Pb , where the deviations from the data on the binding
energies are less than 1.5 MeV. Compared with the previous RHF results, the present DDRHF brings substantial
improvements on these observables.

 \begin{table}[htbp]\centering\setlength{\tabcolsep}{5pt}
\caption{The properties of the nuclei $^{16}$O, $^{40}$Ca, $^{48}$Ca, $^{90}$Zr and $^{208}$Pb predicted by the DDRHF
with PKO1, PKO2, and PKO3, and the RHF approaches with set e in Ref. \cite{Bouyssy:1987}, HFSI  \cite{Bernardos:1993},
ZRL1 \cite{Marcos:2004}.  The total binding energies $E$, the proton spin-orbit splitting $\Delta_{LS}$ for the $1p$
shell  of $^{16}$O and $1d$ shell of $^{40}$Ca and $^{48}$Ca nuclei are given in MeV. The r.m.s. charge radii $r_c$ are
given in fm. The experimental data are taken from Ref. \cite{Audi:1995} for the binding energies, and from Refs.
\cite{Vries:1987, Dutta:1991, Fricke:1995} for charge radii, and from Ref. \cite{Oros:1996} for spin-orbit splitting.
}\label{tab:comparison}
 \begin{tabular}{c|ccc|ccc|ccc|cc|cc}\toprule[1.5pt]\toprule[0.5pt]
&\multicolumn{3}{c|}{$^{16}$O}&\multicolumn{3}{c|}{$^{40}$Ca}&\multicolumn{3}{c|}{$^{48}$Ca}
&\multicolumn{2}{c|}{$^{90}$Zr}&\multicolumn{2}{c}{$^{208}$Pb}\\
Set& $E$&$r_c$&$\Delta_{LS}$&$E$&$r_c$&$\Delta_{LS}$&$E$&$r_c$&$\Delta_{LS}$& $E$&$r_c$& $E$&$r_c$\\
\midrule[0.5pt]
EXP&-127.63&2.693&6.32&-342.05&3.478&5.94&-415.99&3.479&5.01&-783.89&4.270&-1636.45&5.504\\
\hline
PKO1&-128.33&2.676&6.36&-343.28&3.443&6.63&-417.37&3.451&5.59&-784.61&4.251&-1636.92&5.505\\
PKO2&-126.94&2.659&6.25&-340.93&3.427&6.47&-415.55&3.450&6.15&-782.56&4.247&-1636.63&5.509\\
PKO3&-128.34&2.687&6.20&-343.37&3.448&6.63&-416.80&3.459&5.27&-784.89&4.254&-1637.42&5.505\\
\midrule[0.5pt]
Ref.\cite{Bouyssy:1987}&-99.52&2.73&7.3&-280.8&3.47&8.0&-349.44&3.47&4.1&-673.2&4.26&-1406.08&5.47\\
HFSI&-128.64&2.73&6.4&-341.2&3.48&7.05&-414.24&3.48&3.27&-779.4&4.26&-1622.4&5.52\\
ZRL1&-127.68&2.71&6.3&-341.2&3.44&7.1&-417.12&3.49&2.6&-787.5&4.25&-1636.96&5.49\\
\bottomrule[0.5pt]\bottomrule[1.5pt]
\end{tabular}
 \end{table}

We now focus on the comparison of results between the DDRHF with PKO1, PKO2 and PKO3 and the RMF with the non-linear
PK1 \cite{Long04}, NL3 \cite{Lalazissis:1997} and the density-dependent PKDD \cite{Long04}, DD-ME1 \cite{Niksic:2002}.
The reference nuclei are those selected to determine these seven parameterizations, i.e., $^{16}$O, $^{40}$Ca,
$^{48}$Ca, $^{56}$Ni, $^{58}$Ni, $^{68}$Ni, $^{90}$Zr, $^{112}$Sn, $^{116}$Sn, $^{124}$Sn, $^{132}$Sn, $^{182}$Pb,
$^{194}$Pb, $^{204}$Pb, $^{208}$Pb, $^{214}$Pb, and $^{210}$Po. \tabref{tab:Observe} shows the masses (binding
energies) of these nuclei, where the numbers in boldface denote the selected observable used in the corresponding
parameterization. The last row of \tabref{tab:Observe} presents the r.m.s. deviations from the data, defined as
  \beq
\Delta \equiv \lrs{ \sum_i^N\lrb{ E^{\text{EXP}}_i - E^{\text{Cal}}_i}^2/N}^{\ff2}.
  \eeq
It is seen that the DDRHF reproduces well the binding energies of the selected nuclei as compared with the data, where
few (only 4 cases for PKO1) deviations are larger than 1.5 \mev. From the r.m.s. deviations in \tabref{tab:Observe} it
is also demonstrated that the DDRHF achieves successful quantitative descriptions comparable to the RMF, even slightly
better for the selected nuclei. Since the selected nuclei cover from the light to heavy regions, one may conclude that
the DDRHF with PKO series can provide appropriate quantitative descriptions of the masses of finite nuclei throughout
the periodic table.

 \begin{table}[htbp]\centering\setlength{\tabcolsep}{0.15cm}
\caption{The binding energies of the selected nuclei used in the various models. The results are calculated by the
DDRHF with PKO1, PKO2, and PKO3, and the RMF with PK1, PKDD, NL3 and DD-ME1. The experimental data are taken from Ref.
\cite{Audi:1995}. The boldface values correspond to nuclei used in the various fits. }\label{tab:Observe}
 \begin{tabular}{ccccccccc}\toprule[1.5pt]\toprule[0.5pt]
&EXP&PKO1&PKO2&PKO3&PK1&NL3&PKDD&DD-ME1\\ \midrule[0.5pt]
$^{16}$O& -127.619&\bf -128.325&\bf -126.942&\bf-128.345&\bf-128.094&\bf-127.127&\bf-127.808&\bf-127.926\\
$^{40}$Ca&-342.052&\bf-343.275&\bf-340.930&\bf-343.374&\bf-342.773&\bf-341.709&\bf-342.579&\bf-343.653\\
$^{48}$Ca&-415.991&\bf-417.372&\bf-415.554&\bf-416.799&\bf-416.077&\bf-415.377&\bf-415.944&\bf-415.012\\
$^{56}$Ni&-483.998&\bf-483.061&\bf-484.548&\bf-480.807&\bf-483.956&-483.599&\bf-484.479&-480.869\\
$^{58}$Ni&-506.454&-502.679&-503.739&-500.912&-504.033&\bf-503.395&-504.013&-501.312\\
$^{68}$Ni&-590.430&\bf-591.370&\bf-588.581&\bf-590.693&\bf-591.685&-591.456&\bf-591.241&-592.253\\
$^{90}$Zr&-783.893&\bf-784.605&\bf-782.444&\bf-784.891&\bf-784.781&\bf-783.859&\bf-784.879&\bf-784.206\\
$^{112}$Sn&-953.529&-951.651&-950.488&-951.472&-954.210&-952.562&-953.730&\bf-952.468\\
$^{116}$Sn&-988.681&\bf-987.276&\bf-985.808&\bf-987.119&\bf-988.491&\bf-987.699&\bf-988.066&\bf-988.470\\
$^{124}$Sn&-1049.963&-1049.619&-1047.818&-1048.918&-1049.162&\bf-1049.884&-1048.113&\bf-1049.880\\
$^{132}$Sn&-1102.920&\bf-1103.585&\bf-1103.171&\bf-1102.106&\bf-1103.503&\bf-1105.459&\bf-1102.648&\bf-1103.857\\
$^{182}$Pb&-1411.650&\bf-1412.159&\bf-1411.832&\bf-1413.027&-1416.431&-1415.184&-1416.309&-1415.216\\
$^{194}$Pb&-1525.930&\bf-1523.023&\bf-1524.502&\bf-1523.546&\bf-1525.536&-1525.733&\bf-1525.474&-1524.937\\
$^{204}$Pb&-1607.520&-1606.269&-1607.230&-1606.737&-1607.851&-1609.906&-1607.770&\bf-1609.676\\
$^{208}$Pb&-1636.446&\bf-1636.913&\bf-1636.625&\bf-1637.435&\bf-1637.443&\bf-1640.584&\bf-1637.387&\bf-1641.415\\
$^{214}$Pb&-1663.298&\bf-1660.797&\bf-1658.841&\bf-1661.099&\bf-1659.382&\bf-1662.551&\bf-1656.084&\bf-1662.011\\
$^{210}$Po&-1645.228&-1646.350&-1645.877&-1647.242&-1648.443&-1650.755&-1648.039&\bf-1651.482\\
\midrule[0.5pt] $\Delta$&&1.6177&1.8745&2.0489&1.8825&2.2506&2.3620&2.7561\\
\bottomrule[0.5pt]\bottomrule[1.5pt]
\end{tabular}
\end{table}

In the DDRHF parameterizations, the radii are not included as fitted observables. The charge radii of the selected
nuclei calculated by the DDRHF and RMF are shown in \tabref{tab:rch_ob}. As one can see from the r.m.s. deviations, the
DDRHF gives satisfactory quantitative descriptions of the charge radii, less good but still comparable to the RMF
predictions. It is also found that large deviations only exist in the light region. For heavy nuclei, the DDRHF has the
same quality of fit as the RMF. It should also be kept in mind that other corrections beyond the mean field can affect
the radii. For instance, the corrections from the RPA correlations may increase the charge radii for $^{40}$Ca.

 \begin{table}[htbp]\centering\setlength{\tabcolsep}{10pt}
\caption{Same as \tabref{tab:Observe}, for charge radii. The experimental data are taken from Refs. \cite{Vries:1987,
Dutta:1991, Fricke:1995}.}\label{tab:rch_ob}
 \begin{tabular}{ccccccccc}\toprule[1.5pt]\toprule[0.5pt]
&EXP&PKO1&PKO2&PKO3&PK1&NL3&PKDD&DD-ME1\\ \midrule[0.5pt]
$^{16}$O&2.693&2.6763&2.6593&2.6871&2.6957&2.7251&2.6988&2.7268\\
$^{40}$Ca&3.478&3.4428&3.4266&3.4479&3.4433&3.4679&3.4418&3.4622\\
$^{48}$Ca&3.479&3.4506&3.4500&3.4594&3.4675&3.4846&3.4716&3.4946\\
$^{56}$Ni&&3.6903&3.6943&3.6944&3.7085&3.7122&3.7162&3.7315\\
$^{58}$Ni&3.776&3.7199&3.7213&3.7264&3.7383&3.7435&3.7442&3.7613\\
$^{68}$Ni&&3.8501&3.8518&3.8613&3.8621&3.8773&3.8681&3.8926\\
$^{90}$Zr&4.270&4.2505&4.2466&4.2537&4.2522&4.2689&4.2534&4.2725\\
$^{112}$Sn&4.593&4.5635&4.5640&4.5672&4.5704&4.5861&4.5722&4.5901\\
$^{116}$Sn&4.625&4.5930&4.5944&4.5981&4.5984&4.6149&4.6004&4.6212\\
$^{124}$Sn&4.677&4.6481&4.6523&4.6545&4.6536&4.6685&4.6567&4.6781\\
$^{132}$Sn&&4.6967&4.7053&4.7011&4.7064&4.7183&4.7102&4.7270\\
$^{182}$Pb&&5.3710&5.3687&5.3683&5.3708&5.3900&5.3709&5.3896\\
$^{194}$Pb&5.442&5.4333&5.4311&5.4320&5.4327&5.4506&5.4329&5.4539\\
$^{204}$Pb&5.482&5.4856&5.4883&5.4839&5.4869&5.5027&5.4877&5.5038\\
$^{208}$Pb&5.504&5.5054&5.5093&5.5046&5.5048&5.5204&5.5053&5.5224\\
$^{214}$Pb&5.559&5.5633&5.5644&5.5629&5.5658&5.5820&5.5635&5.5779\\
$^{210}$Po&&5.5394&5.5425&5.5383&5.5370&5.5539&5.5371&5.5544\\
\midrule[0.5pt]
$\Delta$&&0.0269&0.0299&0.0225&0.0204&0.0177&0.0188&0.0163\\
\bottomrule[0.5pt]\bottomrule[1.5pt]
\end{tabular}
\end{table}

\section{Conclusions}\label{sec:CP}
In this work, we have introduced an effective Lagrangian suitable for a description of nuclear systems in a
relativistic Hartree-Fock framework. This Lagrangian contains the 4 effective mesons $\sigma, \omega, \rho$ and $\pi$
with density-dependent meson-nucleon couplings. The number of adjustable free parameters is 8, i.e., comparable with
the currently existing Lagrangians where the Fock terms are not treated. We recall that the pion can contribute only
through its Fock term if parity is to be conserved. The criteria for determining the model parameters are the
reproduction of the binding energies in a number of selected nuclei, and the bulk properties of nuclear matter
(saturation point, compression modulus, symmetry energy).

In comparison with other RMF models, the present density-dependent RHF approach is at the same level of quantitative
description, or slightly better as far as the set of selected nuclei are concerned. This applies not only for binding
energies but also for observables not considered in the fitting procedure such as charge radii. Because the nuclei
selected for the parameter fitting cover a wide range of masses, it can be expected that the present density-dependent
RHF approach can perform quite well throughout the periodic table and out to the borderlines of the region of bound
nuclei.

{One may wonder if one has gained any benefit over the RMF approach by including the Fock terms. The answer is twofold.
Firstly, there is less arbitrariness in RHF since one does not drop the exchange terms while they are known to be
important. Even though their effects can be partly simulated by renormalizing the coupling constants, this cannot be
completely correct: for instance, the pion will never appear explicitly in RMF. Secondly, there are other physical
properties not discussed in this paper, where the Fock terms and particularly the pion contributions are quite
important for explaining the nuclear data. This is in particular the case of the isospin dependence of the nucleon
effective mass \cite{Long:2006}, or the evolution of the single-particle energies along isotopic and isotonic chains
\cite{Long:2007}.}

{Several aspects remain to be investigated. The next step is to improve the treatment of pairing correlations within
the RHF by setting up a computational scheme for a relativistic Hartree-Fock-Bogoliubov approach. Another extension is
to include the tensor coupling of the $\rho$-meson and to redetermine the optimal effective Lagrangians for describing
finite nuclei.}

\begin{acknowledgements}
{This work is  partly supported by the National Natural Science Foundation of China under Grant No. 10435010, and
10221003, and the European Community project Asia-Europe Link in Nuclear Physics and Astrophysics CN/Asia-Link
008(94791). }
\end{acknowledgements}

\begin{appendix}
\section{The Microscopic Center-of-Mass Corrections}\label{Sec:CMC}

\subsection{Corrections to the total energy}
The corrections from the center-of-mass motion to the total energy have been systematically
discussed in Ref. \cite{Bender:2000}. Here for completeness, we just recall the corresponding
formula in this subsection. The correction of the center-of-mass motion on the binding energy is
taken into account in a non-relativistic form as,
 \beq
E_\cm = -\frac{1}{2MA} \lc\svec P_\cm^2\rc~,
 \eeq
where the center-of-mass momentum $\svec P_\cm = \sum_i \svec p_i$ and the expectation value of its
square is
 \beq
\lc\svec P_\cm^2\rc = \sum_a v_a^2 p_{aa}^2 - \sum_{a,b} v_a^2 v_b^2
\svec p_{ab}\cdot\svec p_{ab}^* + \sum_{a,b} v_a u_a v_b u_b \svec
p_{ab}\cdot\svec p_{\bar a\bar b}~.
 \eeq
To calculate the above expression, we express $\svec P_\cm^2$ in second quantized form,
 \beq\label{A3}
\svec P_\cm^2 = \sum_{ab} \svec p_a\cdot\svec p_b =\sum_{\alpha\alpha'} \svec p_{\alpha\alpha'}^2 c_\alpha^\dag
c_{\alpha'} +\sum_{\alpha\beta\alpha'\beta'} \svec p_{\alpha\alpha'}\cdot\svec p_{\beta\beta'} c_{\alpha}^\dag
c_\beta^\dag c_{\beta'} c_{\alpha'}~.
 \eeq

In the spherical case, we obtain
 \beq\begin{split}
\lc\hat{\svec P}^2_\cm\rc=&-\sum_{\alpha\kappa} v_{\alpha\kappa}^2\sum_{m=-j}^j \lc \alpha\kappa
m\rl\Delta\lr\alpha\kappa m\rc\\ &-\sum_{\alpha,\beta,\kappa, K}v_{\alpha\kappa} v_{\beta K} \lrb
{v_{\alpha\kappa} v_{\beta K}+u_{\alpha\kappa} u_{\beta K}}\sum_{m=-j}^j\sum_{M=-J}^J\lr\lc
\alpha\kappa m\rl\svec\nabla\lr\beta KM\rc\rl^2~,
 \end{split}\eeq
where
 \beq
\sum_{m=-j}^j\lc \alpha\kappa m\rl\Delta\lr\alpha\kappa m\rc= \hat
j^2\sum_{\eta=\pm}\int drr^2
\psi_{\alpha\kappa}^{(\eta)}\Delta\psi_{\alpha\kappa}^{(\eta)}~,
 \eeq
and
 \beq\begin{split}
\sum_{m=-j}^j\sum_{M=-J}^J\lr\lc \alpha\kappa
m\rl\svec\nabla\lr\beta KM\rc\rl^2=&\hat j^2\hat
J^2\sum_{\eta=\pm}(-1)^{L^{(\eta)}+l^{(\eta)}}
\Lb\bea{ccc}J&j&1\\ l^{(\eta)}&L^{(\eta)}&\ff2\eea\Rb^2\\
&\times \lrs{ L^{(\eta)}\delta_{l^{(\eta)},L^{(\eta)}-1}
A^{(\eta)}_{\alpha\beta}B^{(\eta)}_{\beta\alpha} +l^{(\eta)}\delta_{L^{(\eta)},l^{(\eta)}-1}
A^{(\eta)}_{\beta\alpha}B^{(\eta)}_{\alpha\beta}}~.
 \end{split}\eeq
The corresponding formulas for $A,B$ read as,
 \bsub\beqn
A^{(\eta)}_{\alpha\beta}&=&\int dr r^2\psi_{\alpha\kappa}\lrb{\frac{d}{dr}+\frac{L^{(\eta)}+1}{r}}
\psi_{\beta K}\\
B^{(\eta)}_{\alpha\beta}&=&\int dr r^2\psi_{\alpha\kappa}\lrb{\frac{d}{dr}-\frac{L^{(\eta)}}{r}}
\psi_{\beta K}
 \eeqn\esub
In the above expressions, $\eta$ denotes the corresponding quantities of the upper ($\eta=+$) or
lower ($\eta=-$) components of the Dirac spinor (\ref{SpinorSP}).

\subsection{Corrections to radii}
The definition of the center-of-mass coordinate is
 \beq
\svec r_\cm = \frac{1}{A} \sum_{i=1}^A\svec r_i~.
 \eeq
The nucleon r.m.s. radius is defined as (taking protons as
representative),
 \beq
r_p^2 = \ff Z\lc\phi\rl\sum_{i = 1}^Z \svec r_i^2\lr\phi\rc~.
 \eeq
With the center-of-mass correction, it then becomes
 \beq
r_{\cm,p}^2 =\ff Z\lc\phi\rl\sum_{i = 1}^Z (\svec r_i-\svec r_\cm)^2\lr\phi\rc = r_p^2 + \delta r_p^2, 
 \eeq
where the correction term $\delta r_p^2$ reads as
 \beq \label{A11}
\delta r_p^2 =- \frac{2}{Z}\sum_{i = 1}^Z \lc\phi\rl \svec r_i\cdot\svec r_\cm\lr\phi\rc + \lc\phi\rl \svec
r_\cm^2\lr\phi\rc.
 \eeq
In right hand of above equation, there exist the following type operators
 \beq\label{roperator}
\hat r(i,j) = \sum_{i,j}\svec r_i\cdot\svec r_j = \sum_i \svec r_i^2 + \sum_{i\ne j} \svec r_i\cdot\svec r_j.
 \eeq
Similar as the quantization of the center-of-mass momentum, it can be quantized as
 \beq
\hat r(i,j) = \sum_{\alpha\alpha'} (\svec r^2)_{\alpha\alpha'} c_\alpha^\dag c_{\alpha'} +
\sum_{\alpha\alpha'\beta\beta'} \svec r_{\alpha\alpha'} \cdot\svec r_{\beta\beta'} c_\alpha^\dag c_\beta^\dag
c_{\beta'} c_{\alpha'}~,
 \eeq
With the BCS state,
 \beq
\lr\phi\rc = \lr \text{BCS}\rc = \prod_k \lrb{u_k + v_k c^\dag_k c^\dag_{\bar k}} \lr 0\rc,
 \eeq
the expectation can be obtained as
 \beq
\lc\phi\rl \hat r(i,j) \lr\phi\rc = \sum_\alpha v_\alpha^2 (\svec r^2)_{\alpha\alpha} -\sum_{\alpha\beta} v_\alpha^2
v_\beta^2 \svec r_{\alpha\beta}\cdot\svec r_{\beta \alpha} + \sum_{\alpha\beta}v_\alpha v_\beta u_\alpha u_\beta \svec
r_{\alpha\beta}\cdot\svec r_{\bar \alpha\bar \beta}.
 \eeq
Among the right hand of above equation, the terms $\svec r_{\alpha\beta}\cdot\svec r_{\beta \alpha}$ and $\svec
r_{\alpha\beta}\cdot\svec r_{\bar \alpha\bar \beta}$ can be obtianed as
 \bsub\begin{align}
\svec r_{\alpha\beta}\cdot\svec r_{\beta\alpha} = &\int rdr r'dr' \hat j_a^2\hat j_b^2 \lrs{ \scr A_{\alpha\beta}
G_\alpha G_\beta +\scr B_{\alpha\beta} F_\alpha F_\beta} _r \lrs{ \scr A_{\alpha\beta} G_\alpha G_\beta + \scr
B_{\alpha\beta} F_\alpha F_\beta} _{r'}~,\\
\svec r_{\alpha\beta}\cdot\svec r_{\bar\alpha\bar\beta} =& \int rdr r'dr' \hat j_a^2\hat j_b^2 \lrs{ \scr
A_{\alpha\beta} G_\alpha G_\beta + \scr B_{\alpha\beta} F_\alpha F_\beta} _r \lrs{ \bar {\scr A}_{\alpha\beta} G_\alpha
G_\beta +\bar {\scr B}_{\alpha\beta} F_\alpha F_\beta} _{r'}~,
 \end{align}\esub
where
 \bsub \begin{align}
\scr A_{\alpha\beta} =&(-1)^{l_a}\hat l_b \Lb\bea{ccc}j_b&j_a&1\\
l_a&l_b&\ff2\eea\Rb C_{l_b010}^{l_a0}& \scr B_{\alpha\beta} =
(-1)^{l_a'}\hat l_b'\Lb\bea{ccc}j_b&j_a&1\\
l_a'&l_b'&\ff2\eea\Rb C_{l_b'010}^{l_a'0}~,\\
\bar{\scr  A}_{\alpha\beta} =&(-1)^{l_b}\hat l_b \Lb\bea{ccc}j_b&j_a&1\\
l_a&l_b&\ff2\eea\Rb C_{l_b010}^{l_a0}&\bar{\scr  B}_{\alpha\beta} = (-1)^{l_b'}\hat l_b'\Lb\bea{ccc}j_b&j_a&1\\
l_a'&l_b'&\ff2\eea\Rb C_{l_b'010}^{l_a'0}~.
 \end{align}\esub

In the end, the correction terms (\ref{A11}) for the proton and neutron radii can be expressed as,
 \bsub\begin{align}
\delta r_p^2 =& -\frac{1}{A}\lrs{\lrb{  2\bar r_p^2 -\bar r_M^2} + \lrb{  2 \dbar r_p^2 - \dbar r_M^2}}~, & 
\delta r_n^2 =& -\frac{1}{A}\lrs{\lrb{2\bar r_n^2 -\bar r_M^2} + \lrb{2 \dbar r_n^2 - \dbar r_M^2}}~, 
 \end{align}\esub
where
 \bsub\label{CMC:radii}\begin{align}
\bar r_p^2 =& \ff Z\int r^4 \rho_b^{(p)}dr,&\dbar r_p^2=& -\ff Z\sum_{\alpha\beta}^Z v_\alpha v_\beta\lrb{   v_\alpha
v_\beta\svec r_{\alpha\beta}\cdot\svec r_{\beta\alpha}- u_\alpha u_\beta \svec
r_{\alpha\beta}\cdot\svec r_{\bar \alpha\bar \beta}},\\
\bar r_n^2 =& \ff N\int r^4 \rho_b^{(n)}dr,& \dbar r_n^2=& -\ff N\sum_{\alpha\beta}^N v_\alpha v_\beta\lrb{   v_\alpha
v_\beta\svec r_{\alpha\beta}\cdot\svec r_{\beta\alpha}- u_\alpha u_\beta \svec
r_{\alpha\beta}\cdot\svec r_{\bar \alpha\bar \beta}},\\
\bar r_M^2 =& \ff A\lrb{   Z\bar r_p^2 + N\bar r_n^2} ,& \dbar r_M^2=&+\ff A\lrb{   Z \dbar r_p^2 + N\dbar r_n^2}.
 \end{align}\esub

\subsection{Corrections to the charge density}
In general, the charge density is obtained from the calculated
proton distribution corrected by the center-of-mass motion and
finite proton size \cite{Negele:1970, Campi:1972}. The first
correction is done by using the proton density $\rho_\cm$ in the
center-of-mass system, which is related to the HF density through
 \beq\label{density_CM}
\rho_{\text{HF}} = \frac{4}{B^3\pi^{\ff2}}\int e^{-r'^2/B^2}
\rho_\cm(\lr \svec r-\svec r'\rl) d\svec r'~,
 \eeq
with $B = \lrs{\hbar /M\omega A}^{\ff2}$. This formula is valid for a system with the
harmonic-oscillator center-of-mass correction, which gives the center-of-mass correction on the
energy as
 \beq
E_\cm = -\frac{3}{4}\hbar \omega = -\frac{1}{2M A}\lc\svec
P_\cm^2\rc~.
 \eeq
In term of the center-of-mass momentum, the factor $B$ can be written as,
 \beq
B^{-2} = {\hbar^{-1}} {M A\omega }= \hbar ^{-2}{MA\hbar\omega}
=\frac{2}{3} \hbar ^{-2}{\lc\svec P_\cm^2\rc}~.
 \eeq
The HF density (\ref{density_CM}) is then expressed as
 \beq
\rho_{\text{HF}} = \frac{8}{3}\sqrt{\frac{2\lc\svec P_\cm^2\rc^3}{3\pi}} \int
\exp\lrs{-\frac{2}{3}\lc\svec P_\cm^2\rc r'^2}\rho_\cm(\lr \svec r-\svec r'\rl) d\svec r'~.
 \eeq
Such that, the contributions from the center-of-mass motion to the density can be treated in microscopic way by
introducing the expectation of the center-of-mass momentum (\ref{A3}).

The finite proton is also taken into account by convoluting $\rho_\cm$ with a Gaussian representing the proton form
factor,
 \beq\label{chargedendis1}
\rho_{\text{ch}}(r) = \frac{1}{2\pi^2 r}\int_0^\infty k\sin(kr) \bar \rho_{\text{HF}}(k)
\exp\lrs{\ff4 k^2(B^2 - a^2)} dk~,
 \eeq
where the Gaussian is $\lrb{   a\sqrt{\pi}}^{-3} \exp(-r^2/a^2)$, with $a = \sqrt{2/3} \lc
r_p\rc_{\text{r.m.s.}}$, $\lc r_p\rc_{\text{r.m.s.}} = 0.8$ fm and $\bar \rho_{\text{HF}}(k)$ is
the Fourier transform of the HF proton density. When $a> B$, this expression can be reduced into
the following form,
 \beq\begin{split}
\rho_{\text{ch}}(r)=& \int d\svec r_2 \lrb{   a\sqrt \pi}^{-3}\exp\lrs{-(\svec
r-\svec r_2)^2/a^2}\rho_\cm(r)\\
=&\int d\svec r_1d \svec r_2\lrb{   a\sqrt \pi}^{-3}\lrb{   B\sqrt \pi}^{-3}\exp\lrs{ -(\svec r
-\svec r_2)^2/a^2 + (\svec r_2-\svec r_1)^2/B^2}
\rho_{\text{HF}}(r_1)\\
=&\int d\svec r_1\lrs{\lrb{   a^2 - B^2}\pi}^{-3/2} \exp\lrs{-(\svec r-\svec r_1)^2/\lrb{   a^2 -
B^2}} \rho_{\text{HF}}(r_1)~.
 \end{split}\eeq
Denoting by $\lambda^2 =1/( a^2 -B^2) $, the integral can be
expressed as,
 \beq
\rho_{\text{ch}}(r) = \frac{\lambda^3}{\pi^{3/2}}\int \exp\lrs{ -\lambda^2(\svec r-\svec r')^2}
\rho_{\text{HF}}(r') d\svec r'~.
 \eeq
In the spherical case, the integral reads as
 \beq\begin{split}
\rho_{\text{ch}}(r) = &\frac{\lambda^3}{\pi^{3/2}}\int \exp\lrs{ -\lambda^2(\svec
r-\svec r')^2} \rho_{\text{HF}}(r') d\svec r'\\
=&\frac{\lambda^3}{\pi^{3/2}}\int r'^2 dr'\rho_{\text{HF}}(r') \int_0^\pi \sin\vartheta d\vartheta
\exp\lrs{ -\lambda^2(\svec r-\svec r')^2}
\int_0^{2\pi} d\varphi\\
=&\frac{\lambda^3}{\pi^{3/2}}\int r'^2 dr'\rho_{\text{HF}}(r')\exp\lrs{-\lambda^2(r^2 + r'^2)}
\int_0^\pi
\sin\vartheta d\vartheta \exp(2 \lambda^2 rr'\cos\vartheta)\times 2\pi\\
=&\frac{\lambda^3}{\pi^{3/2}}\int r'^2 dr'\rho_{\text{HF}}(r')\exp\lrs{-\lambda^2(r^2 + r'^2)}
\frac{e^{2\lambda^2rr'} - e^{-2\lambda^2 rr'}}{2\lambda^2 rr'}\times 2\pi~.
 \end{split}\eeq
Finally, one can write the charge density distribution combined with the center-of-mass motion and proton size
corrections as,
 \beq
\rho_{\text{ch}}(r) = \frac{\lambda}{\sqrt{\pi r^2}}\int r'dr'\rho_{\text{HF}}(r')\lrs{
e^{-\lambda^2(r-r')^2} - e^{-\lambda^2 (r+r')^2} }~.
 \eeq
\end{appendix}

%

\end{document}